\documentclass[11pt,a4paper]{article}
\usepackage{jheppub}
\usepackage{color}
\usepackage{subfigure}

\newcommand{\be}{\begin{equation}}
\newcommand{\bea}{\begin{eqnarray}}
\newcommand{\eea}{\end{eqnarray}}
\newcommand{\ba}{\begin{array}}
\newcommand{\ea}{\end{array}}
\newcommand{\ee}{\end{equation}}

\def\bse{\begin{subequations}}
\def\ese{\end{subequations}}

\title{Holographic p-wave Josephson junction}

\author{Yong-Qiang Wang \footnote{Corresponding author.},}
\author{Yu-Xiao Liu}
\author{and Zhen-Hua Zhao}

\affiliation{Institute of Theoretical Physics, Lanzhou University,
Lanzhou 730000, People¡¯s Republic of China}
\emailAdd{yqwang@lzu.edu.cn, liuyx@lzu.edu.cn, zhaozhh02@gmail.com}

\abstract{
In this work we generalized holographic model for s-wave DC Josephson junction constructed in arXiv:1101.3326[hep-th]
 to a holographic description for p-wave Josephson junction.  By solving numerically the
coupled equations of motion of Yang-Mills theory for a non-Abelian $SU(2)$ gauge fields in (3+1)-dimensional AdS spacetimes,  we shown that DC current of the p-wave  Josephson junction is proportional to the sine of the phase difference across the junction like  the s-wave case.
}


\begin{document}

\maketitle

\section{Introduction}
The AdS/CFT correspondence~\cite{Maldacena:1997re} has attracted  much attention over the
past few years. The AdS/CFT correspondence relates a gravity theory in a weakly curved
(d + 1)-dimensional AdS spacetimes to the strong coupled d-dimensional field theory living at the AdS boundary.
 In the last few years, with the application of the AdS/CFT correspondence to the superconducting phase, ones find that black holes with charged scalar hair in AdS spacetimes can provide a holographically dual description
of superconductivity~\cite{Gubser:2008px,Hartnoll:2008vx,Hartnoll:2008kx}.  For reviews on
holographic superconductors, see~\cite{Hartnoll:2009sz,Herzog:2009xv,Horowitz:2010gk}.

Recently, a holographic model for s-wave  Josephson junction with DC current constructed in~\cite{Horowitz:2011dz}. Solving numerically the equations of motion of Einstein-Maxwell-scalar model with  spatial dependence $\mu$ and non-vanishing constant $J$, ones show that the DC  current of s-wave Josephson junction is proportional to the sine of the phase difference across the junction. Meanwhile, they also studied dependence of the maximum current on the temperature and size of the junction, which matches precisely with the results for condensed matter physics. This holographic model is extended to  (3+1)-dimensional Josephson junction  in \cite{Wang:2011rva,Siani:2011uj}. Another different
way to construct a holographic Josephson junction based on designer multigravity is studied in~\cite{Kiritsis:2011zq}.

As is well known, in condensed matter systems, SNS Josephson junction can be  made up of two p-wave superconductors. A holographic description of p-wave superconductors has been developed in~\cite{Gubser:2008zu,Gubser:2008wv,Roberts:2008ns}. Naturally,
it would also be of interest to set up a holographic model for p-wave  Josephson junction. 
In this paper, we would like to generalize the
work \cite{Horowitz:2011dz} to a holographic description for the  p-wave  Josephson junction.  We will study a non-Abelian $SU(2)$ Yang-Mills theory in (3+1)-dimensional AdS spacetimes  and  solve numerically  the
coupled, nonlinear partial differential equations of motion,
and analyze the phase dependence of the Josephson current between two p-wave superconductors.

The paper is organized as follows. In Sec. \ref{sec2}, we
review  non-Abelian $SU(2)$ Yang-Mills theory in (3+1)-dimensional AdS spacetimes  and  set up a gravity dual of a (2+1)-dimensional p-wave Josephson junction.
In Sec. \ref{sec3}, we show numerical results of the EOMs
and study the characteristics of the (2+1)-dimensional holographic p-wave Josephson junctions. The last section is devoted to conclusion.

 \section{Holographic setup of p-wave Josephson junction}\label{sec2}
Now let us concentrate on the dynamics of Yang-Mills theory for a non-Abelian $SU(2)$ gauge fields in (3+1)-dimensional AdS spacetimes.
First, we write 
the action of (3+1)-dimensional AdS gravity as
\be \label{graction}
S=\int d^4x \sqrt{-g}(R+\frac{6}{L^2}) \;,
\ee
where $L$ is the curvature radius of asymptotic AdS Spacetimes.
The planar Schwarzschild black hole can be written as:
\be\label{matrix}
ds^2=-f(r)dt^2+\frac{dr^2}{f(r)}+r^2(dx^2+dy^2)\;,\qquad f(r)=\frac{r^2}{L^2}-\frac{M}{r}\;,
\ee
where $M$ is the mass of the black hole.
 The Hawking  temperature of the black hole reads as
\be
T \equiv
\frac{1}{4 \pi } \frac{df}{dr}\bigg|_{r=r_H}=\frac{3}{4\pi}\frac{M^{1/3}}{ L^{4/3}},
\ee
where  $ r_{H}=M^{\frac{1}{3}}L^{\frac{2}{3}}$ is the event horizon  of the black hole.
In this background of gravity, we
now consider a non-Abelian $SU(2)$ gauge fields, with
the action
\begin{eqnarray}
S=\int d^4x\sqrt{-g}\left(
-\frac{1}{4} F^{a\mu\nu}F^a_{\mu\nu}
\right) ,
\end{eqnarray}
where $F_{\mu\nu}^a\equiv\partial_\mu A_\nu^a-\partial_\nu A_\mu^a+\epsilon^{abc}A_\mu^b A_\nu^c$ is the field strength of the SU(2) gauge theory,
and one form $A=A_{\mu} dx^{\mu}=A_{\mu}^a\tau^a dx^{\mu}$ is gauge field,  $\tau^a$ are the generators of $SU(2)$
with $a = 1,2,3$.

Variation of the action (\ref{graction}) with respect to the gauge field $A_{\mu}$ leads to the 
equations of motion
\be\label{max}
D^\mu F_{\mu\nu} =  0
\,,
\ee
where  the gauge covariant derivative is  $D_\mu\equiv\nabla_\mu+i[A_\mu,\,\,\,\,]$.

In order to obtain a direction along
which the SNS stack is arranged,
we take the gauge field ansatz as
\be  \label{ansz}
 A=\phi(r,y)\tau^3\;dt+w(r,y) \;\tau^1 dx+A_r(r,y)\;\tau^3\;dr+A_y(r,y)\;\tau^3\;dy\;,
\ee
where $\phi$, $w$, $A_r$, and $A_y$ are real functions of $r$ and $y$.  Thus the junction can be along the $y$ direction.

With the black hole background (\ref{matrix}) and the above ansatz (\ref{ansz}), the equations of  SU(2) gauge theory (\ref{max}) can be  written as:
\begin{align}
\partial_{r}^2\,w+\frac{1}{r^2f}\partial_{yy}^2\,w+\frac{f'}{f}\partial_r \,w+\left(\frac{\phi^2}{f^2}-A_r^2-\frac{A_y^2}{r^2 f}\right)|\psi|&=0\;,\label{psi}\\
\partial_{r}^2\,\phi+\frac{1}{r^2f}\partial_{y}^2\,\phi+\frac{2}{r}\partial_r \phi-\frac{w^2}{r^2 f}\phi &=0\;,\label{phi}\\
\partial_{y}^2\,A_r-\partial_{r}\partial_y \,A_y-w^2 A_r&=0\;,\label{Ar}\\
\partial_{r}^2\,A_y-\partial_{r}\partial_y\,A_r+\frac{f'}{f}\left(\partial_r A_y-\partial_y A_r\right)-\frac{w^2}{r^2 f}A_y&=0\;,\label{Ay}\\
\partial_{r}A_r+\frac{1}{r^2f}\partial_y\,A_y+\frac{2}{w}\left(A_r\partial_r w+\frac{A_y}{r^2\,f}\partial_y w\right)+\frac{f'}{f}A_r&=0\;.\label{A}
\end{align}
where a prime denotes derivative with respect to $r$.
Because Eqs.~(\ref{psi})-(\ref{A}) are coupled nonlinear equations, one can not solve these equations
analytically. However, it is straightforward to solve them numerically.

In order to solve these coupled equations, first, we need to impose regularity at the horizon and the boundary on the radial coordinate.
Near the horizon ($r=r_H$), the field $\phi$ should be regular:
\be
\phi(r_H)=0
\ee
Near the AdS boundary ($r\rightarrow\infty$),
the Yang-Mills fields take the asymptotic forms
\begin{align}
w&=w^{(1)}(y)+\frac{w^{(2)}(y)}{r}
+\mathcal{O}\left(\frac{1}{r^2}\right)\;,\\
\phi &=\mu(y)-\frac{\rho(y)}{r}+\mathcal{O}\left(\frac{1}{r^2}\right)\;,\\
A_r&=\mathcal{O}\left(\frac{1}{r^3}\right)\;,\\
A_y&=\nu(y)+\frac{J}{r}+\mathcal{O}\left(\frac{1}{r^2}\right)\; \label{defnu},
\end{align}
here, $\mu$ is the chemical potential and  $\rho$ is the charge density,  $\nu$ and $J$  are the superfluid velocity and current, respectively~\cite{Basu:2008bh,Zeng:2010fs}.
It is worth pointing out that current $J$ is a constant and along the $y$ direction. 
For $w^{(1)}$ and $w^{(2)}$ are normalizable, one can impose the condition either $w^{(1)}$ or $w^{(2)}$ vanishes. For simplicity, we will adopt the constraint $w^{(1)}=0$ and interpret $\langle\mathcal{O}\rangle=w^{(2)}$  as the condensate.  Moreover, we introduce the  phase difference  $\gamma =  -\int A_y dy$, which is the gauge invariant and can be rewritten as
\be
\gamma=- \int_{-\infty}^\infty dy \;[\nu(y) -\nu(\pm\infty)]\;.
\label{gamma}
\ee

Furthermore,
we impose the Dirichlet-like boundary condition on $A_r$ and  Neumann-like boundary condition on $w$, $\phi$ and $A_y$ at  the spatial coordinate $y=0$.
At $y=\pm\infty$, the field functions are $y$-independent. Thus the boundary conditions of the coupled equations (\ref{psi})-(\ref{A}) are determined by $J$ and $\mu$.

Ones need  introduce the critical temperature $T_c$ of the p-wave junction, which is proportional to $\mu(\infty)=\mu(-\infty)$ :
\be\label{T_c}
T_c=\frac{3}{4\pi}\frac{\mu(\infty)}{\mu_c}\;,
\ee
where $\mu_c\approx 3.65$. The effective critical temperature inside the gap can be written as
\be\label{T_0}
T_0=\frac{3}{4\pi}\frac{\mu(0)}{\mu_c}\;.
\ee
In order to describe an SNS Josephen junction, we  also choose the  below  $\mu(y)$  which is the same as that in \cite{Horowitz:2011dz}:
\be\label{profile}
\mu(y)=\mu_\infty\left\{1-\frac{1-\epsilon}{2\tanh(\frac{L}{2\sigma})}\left[\tanh\left(\frac{y+\tfrac{L}{2}}{\sigma}\right)-\tanh\left(\frac{y-\tfrac{L}{2}}{\sigma}\right)\right]\right\}\;,
\ee
where the chemical potential $\mu$ is proportional to $\mu_{\infty}\equiv\mu(\infty)=\mu(-\infty)$ at $y=\pm\infty$, and $L$ is the width of p-wave junction. The parameters $\sigma$ and $\epsilon$ control the steepness and depth of p-wave junction, respectively.

\section{Numerical results}\label{sec3}

\begin{figure*}
\begin{center}
 \subfigure[$\phi$]{
  \includegraphics[width=0.4\textwidth]{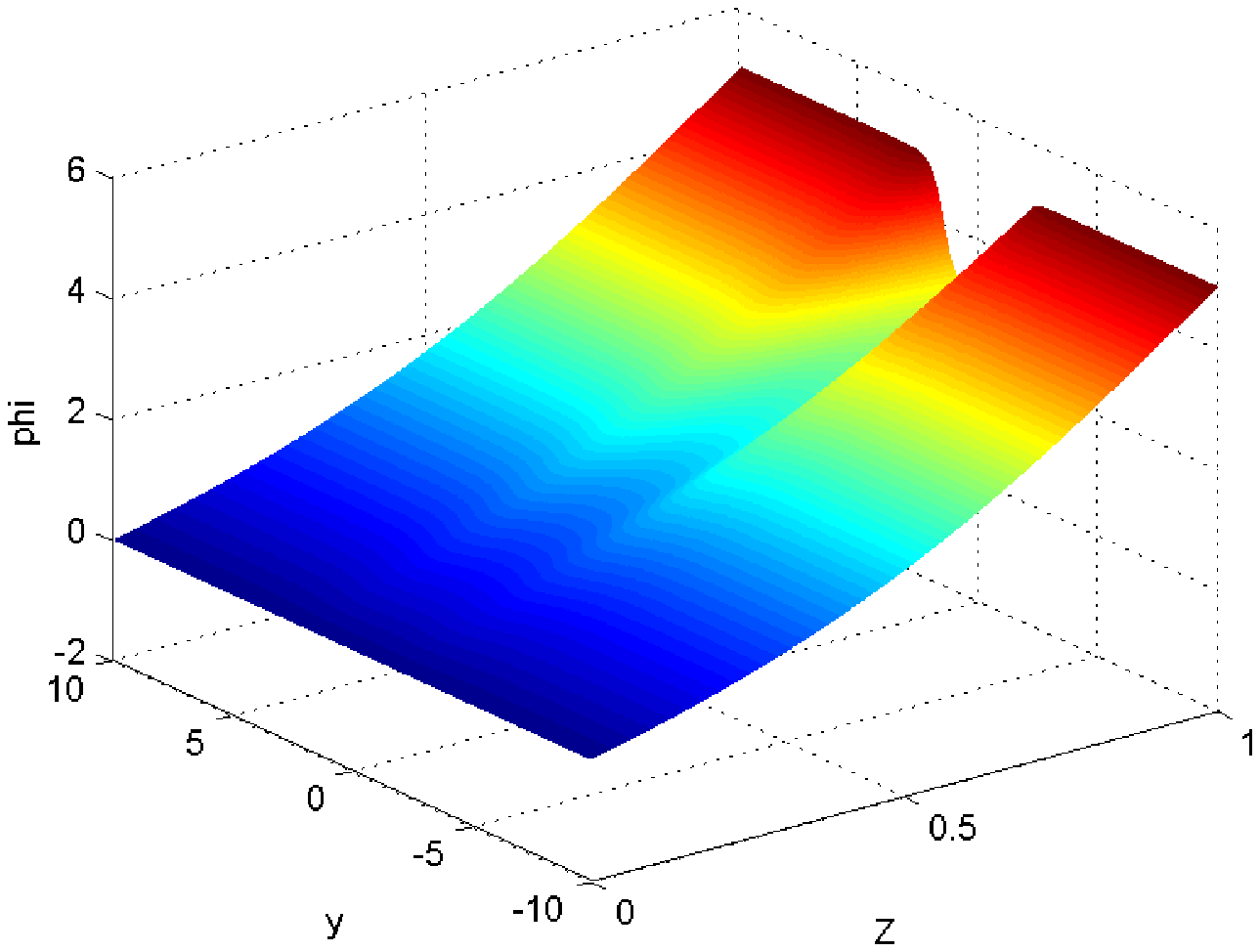}}
\hspace{1.0cm}
 \subfigure[$A_y$]{
  \includegraphics[width=0.4\textwidth]{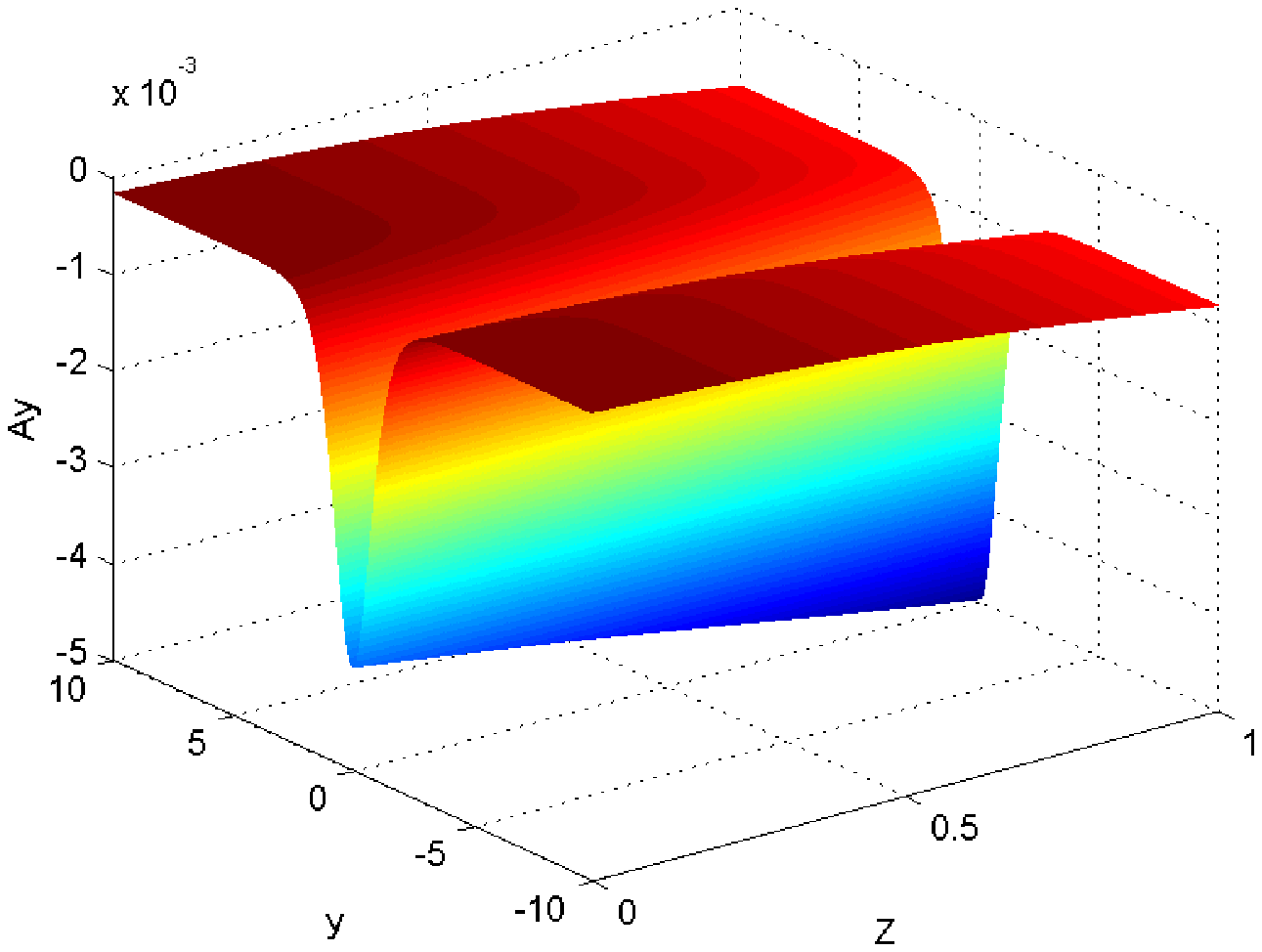}}
\end{center}  \vskip -5mm
\caption{The components  $\phi$ and  $A_y$ of Yang-Mills fields. The parameters are
set to $J/T^2_c=0.0094$, $\mu_\infty=5$, $L=3$, $\epsilon=0.6$, $\sigma=0.5$.}
 \label{fig_scalar_warpfactor}
\end{figure*}

In this section, we will solve the coupled  equations (\ref{psi})-(\ref{A}) numerically with spectral methods.  In order to solve EOMs numerically , we need set the change of variables
$z=1-r_H/r$ and $\tilde y=\tanh(\frac{y}{4\sigma})$. First, the graphs for $\phi$ and $A_y$  with $J/T^2_c=0.0094$,$\mu_\infty=5$, $L=4$, $\epsilon=0.6$, and $\sigma=0.5$, are shown in Fig. \ref{fig_scalar_warpfactor}.

\begin{figure}
  \begin{center}
  \includegraphics[width=0.7\textwidth]{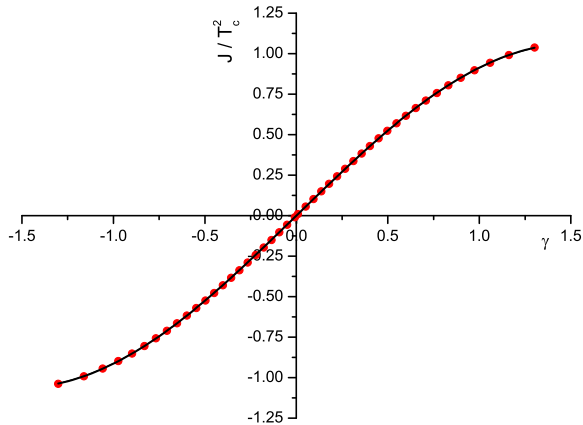}
  \end{center}
  \caption{Superfluid current $J_{\max}/T^2_c$ as the function of the phase difference $\gamma$. The black line is  the sine curve. The parameters are
set to $\mu_\infty=5$, $L=3$, $\epsilon=0.6$, $\sigma=0.5$.}\label{sine}
\end{figure}

Then, in Fig.~\ref{sine}, we show that the DC current of p-wave junction is proportional to the sine of the phase difference across the junction, that is to say, the red dots coming from the numerical calculations match with the the black solid sine curve. Fitting the sine curve to the data, we can obtain the maximum current across the junction:
 \be
 J_{\max}/T^2_c \approx 1.078.
 \ee

\begin{figure*}
  \begin{center}
  \subfigure[$$]{{\label{figphi}}
  \includegraphics[width=0.7\textwidth]{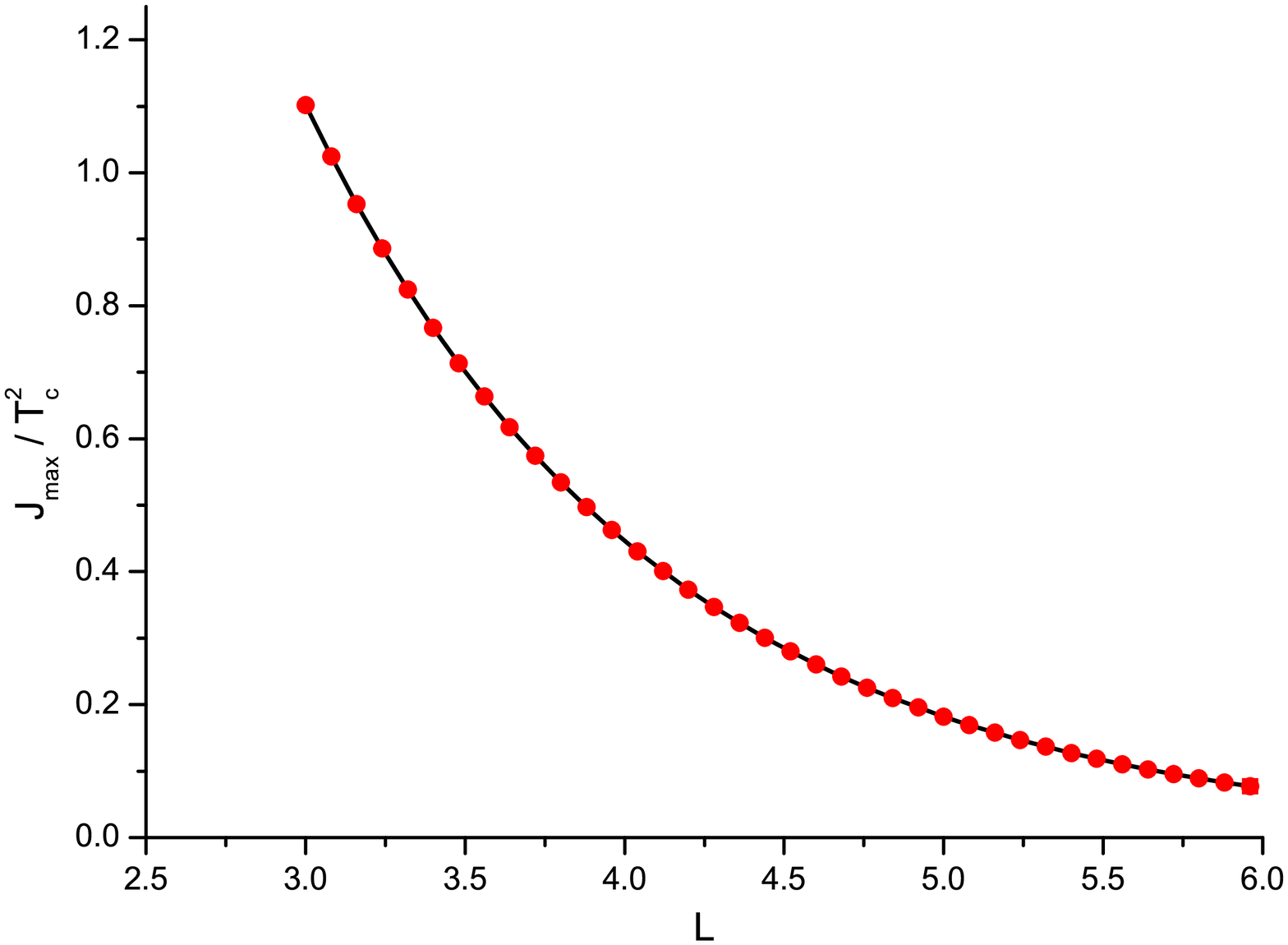}}
 \subfigure[$$]{{\label{figphiprime}}
  \includegraphics[width=0.7\textwidth]{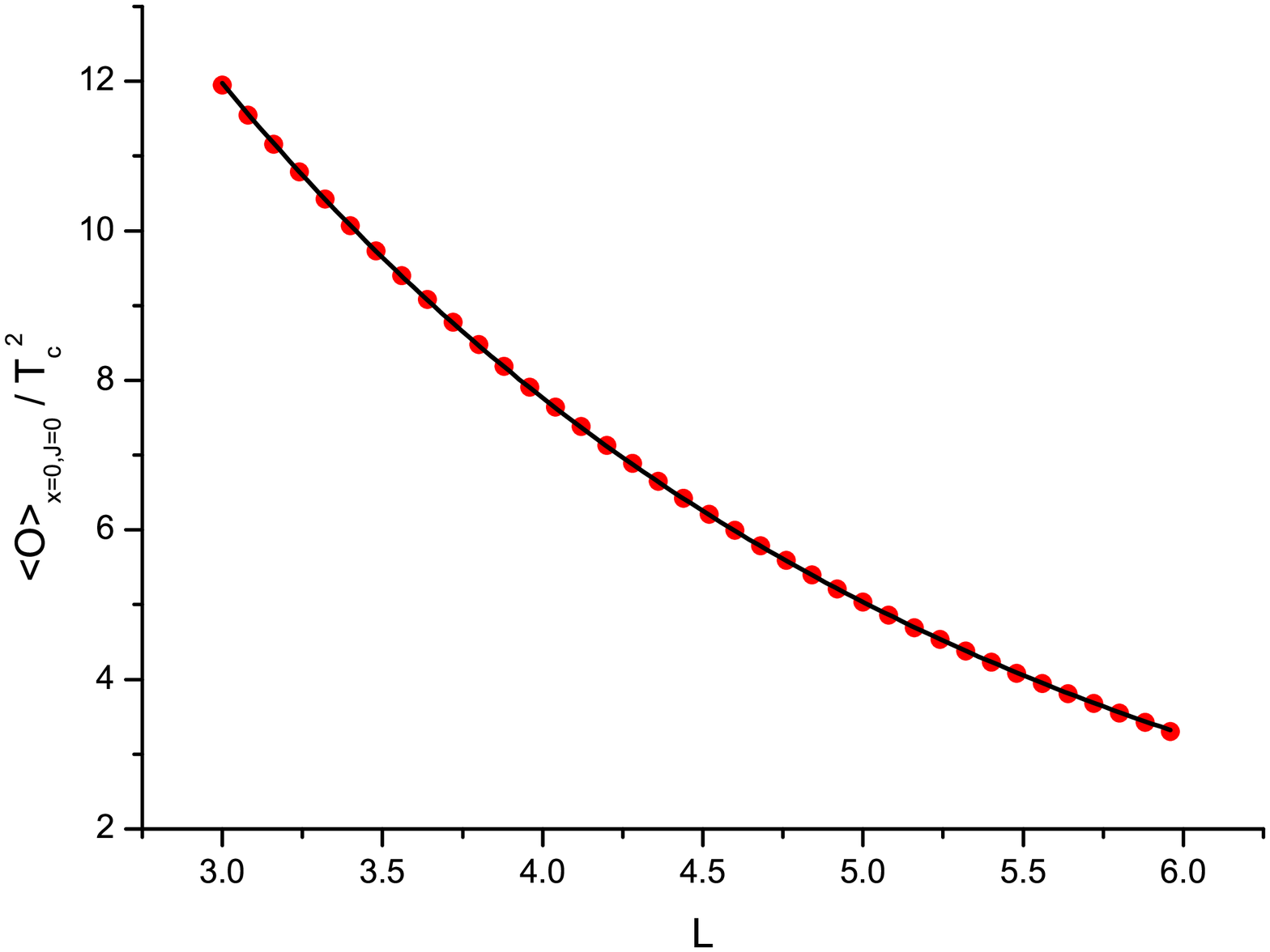}}
  \end{center}
\caption{The curve of $J_{\max}$  and  $\langle\mathcal{O}\rangle_{x=0}$ on $L$. The parameters are
set to $\mu_\infty=5$, $\epsilon=0.6$, $\sigma=0.5$.}
 \label{fig2}
\end{figure*}

\begin{figure}
  \begin{center}
  \includegraphics[width=0.7\textwidth]{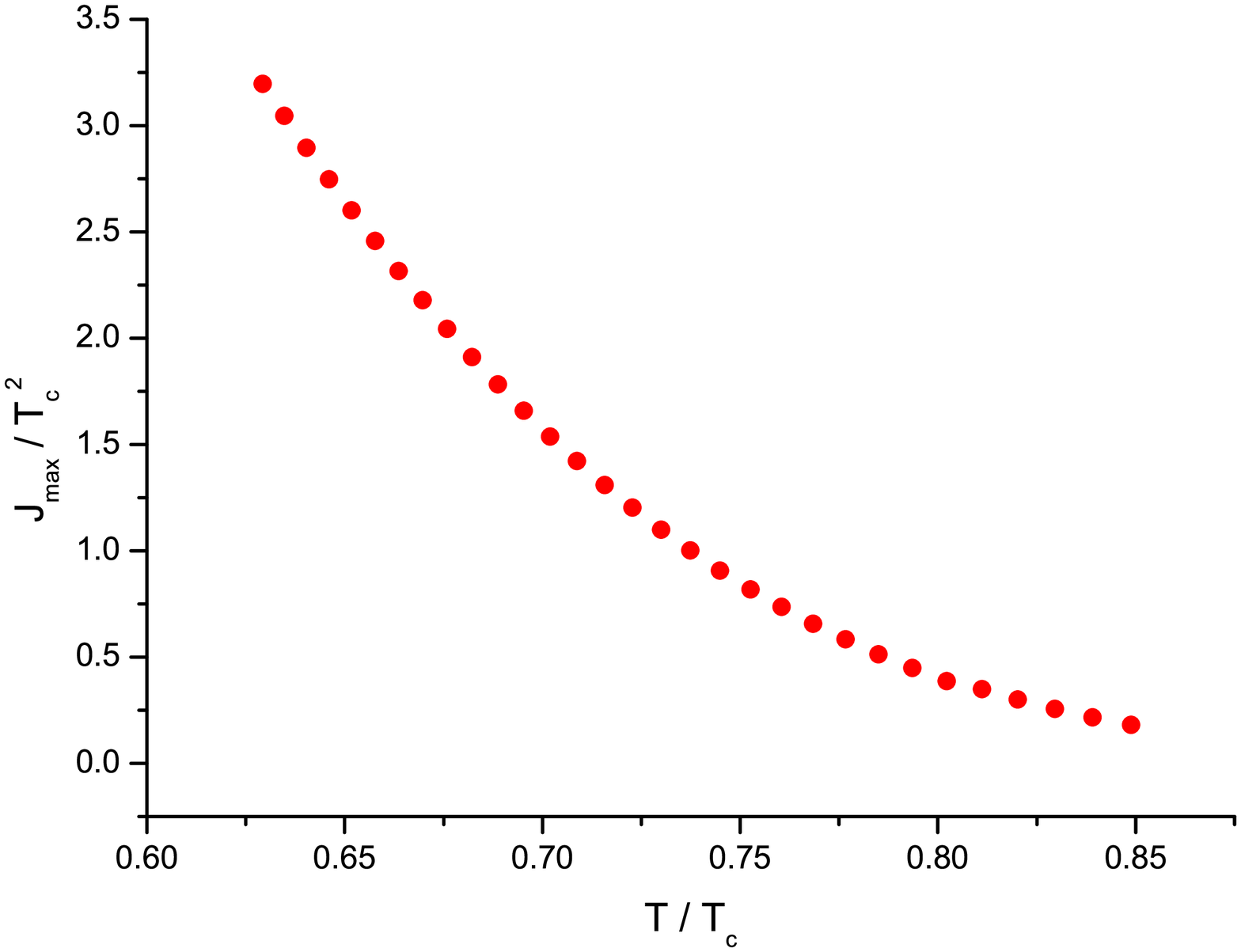}
  \end{center}
  \caption{The curve of $J_{\max}$ on $T$, The parameters are
set to $\epsilon=0.6$, $\sigma=0.5$}\label{ffff}
\end{figure}

The dependence of $J_{\max}$  on the width of the gap is shown in Fig.~\ref{figphi}. The graph predicts an exponential decay with the growing width of the gap in $J_{\max}$:
\be
\frac{J_{\max}}{T^2_c} = A_0\,e^{-\frac{\ell}{\xi}},\label{jmaxell}
\ee
where $\xi$ is the coherence length.
In Fig.~\ref{figphiprime}, the condensate $\langle\mathcal{O}\rangle=\psi^{(2)}$  at zero current is shown. The graph also predicts an exponential decay with the growing width of the gap in $J_{\max}$:
\be
\qquad \frac{\langle \mathcal{O}(0)\rangle}{T^2_c}|_{J=0}= A_1\,e^{-\frac{\ell}{2\,\xi}}.\label{oell}
\ee
Fitting Eq.~(\ref{jmaxell}) and Eq.~(\ref{oell}) with the two sets of data, we can obtain $\{\xi ,A_0 \}  \approx \{1.11,16.42\}$  and $\{\xi,A_1\} \approx \{1.15,43.89\}$ for Eq.~(\ref{jmaxell}) and Eq.~(\ref{oell}), respectively.  The disagreement of two values of $\xi$ is about 4 percent.
In Fig.~\ref{ffff}, we obtain the relation of $J_{\max}$ and $T$.  One can verify that near the critical temperature $T_c$, $J_{\max}$ can reach zero. Since $\epsilon=0.6$, we show the region corresponds to $T/T_c<0.6$, which depicts the character of a p-wave Josephson junction.

\section{Conclusion}

In this paper, we construct a holographic model for p-wave SNS Josephson junction with DC current.  In the background of (3+1)-dimensional gravity, we solve a set of couple, partial differential equations of  a SU(2) gauge field
numerically.  By choosing spatial y-dependence $\mu$, we find the the DC current of p-wave junction is proportional to the sine of the phase difference across the junction.
Moreover, the graph which predicts an exponential decay with the growing width of gap in $J_{\max}$ is obtained. At last,
the curve of $J_{\max}$ and $T$ is also obtained. 
One can see that the model
of the holographic p-wave Josephson junction in our work can match precisely with the results for condensed matter physics. In future, It would be interested for us to
extend to investigate the holographic d-wave Josephson junction.

\section*{Acknowledgement}
This work was supported in part by the National Natural Science Foundation of China (No. 11005054 and No. 11075065), and the
Fundamental Research Fund for Physics and Mathematics of Lanzhou University (No. LZULL200912).  Z.H. Zhao was supported by the Scholarship Award for Excellent Doctoral Student granted by Ministry of Education.


\begin{thebibliography}{99}

\bibitem{Maldacena:1997re}
  J.~M.~Maldacena,
  {\em The Large $N$ limit of superconformal field theories and supergravity},
  Adv.\ Theor.\ Math.\ Phys.\  {\bf 2}, 231 (1998)
  [Int.\ J.\ Theor.\ Phys.\  {\bf 38}, 1113 (1999)]
  [arXiv:hep-th/9711200].

\bibitem{Gubser:2008px}
  S.~S.~Gubser,
  {\em Breaking an Abelian gauge symmetry near a black hole horizon},
  Phys.\ Rev.\  {\bf D 78}, 065034 (2008)
  [arXiv:0801.2977 [hep-th]].

\bibitem{Hartnoll:2008vx}
  S.~A.~Hartnoll, C.~P.~Herzog and G.~T.~Horowitz,
  {\em Building a Holographic Superconductor},
  Phys.\ Rev.\ Lett.\  {\bf 101}, 031601 (2008)
  [arXiv:0803.3295 [hep-th]].

\bibitem{Hartnoll:2008kx}
  S.~A.~Hartnoll, C.~P.~Herzog and G.~T.~Horowitz,
  {\em Holographic Superconductors},
  JHEP {\bf 0812}, 015 (2008)
  [arXiv:0810.1563 [hep-th]].


\bibitem{Hartnoll:2009sz}
  S.~A.~Hartnoll,
  {\em Lectures on holographic methods for condensed matter physics},
  Class.\ Quant.\ Grav.\  {\bf 26}, 224002 (2009)
  [arXiv:0903.3246 [hep-th]].



\bibitem{Herzog:2009xv}
  C.~P.~Herzog,
  {\em Lectures on Holographic Superfluidity and Superconductivity},
  J.\ Phys.\ {\bf A  42}, 343001 (2009)
  [arXiv:0904.1975 [hep-th]].

\bibitem{Horowitz:2010gk}
  G.~T.~Horowitz,
  {\em Introduction to Holographic Superconductors},
  arXiv:1002.1722 [hep-th].


\bibitem{Horowitz:2011dz}
  G.~T.~Horowitz, J.~E.~Santos and B.~Way,
  {\em A Holographic Josephson Junction},
  arXiv:1101.3326 [hep-th].

\bibitem{Wang:2011rva}
  Y.~Q.~Wang, Y.~X.~Liu and Z.~H.~Zhao,
  {\em Holographic Josephson Junction in 3+1 dimensions},
  arXiv:1104.4303 [hep-th].





\bibitem{Siani:2011uj}
  M.~Siani,
  {\em On inhomogeneous holographic superconductors},
  arXiv:1104.4463 [hep-th].

\bibitem{Kiritsis:2011zq}
  E.~Kiritsis and V.~Niarchos,
  {\em Josephson Junctions and AdS/CFT Networks},
  JHEP {\bf 1107}, 112 (2011)
  [arXiv:1105.6100 [hep-th]].





\bibitem{Gubser:2008zu}
  S.~S.~Gubser,
  {\em Colorful horizons with charge in anti-de Sitter space},
  Phys.\ Rev.\ Lett.\  {\bf 101}, 191601 (2008)
  [arXiv:0803.3483 [hep-th]].




\bibitem{Gubser:2008wv}
  S.~S.~Gubser and S.~S.~Pufu,
  {\em The Gravity dual of a p-wave superconductor},
  JHEP {\bf 0811}, 033 (2008)
  [arXiv:0805.2960 [hep-th]].

\bibitem{Roberts:2008ns}
  M.~M.~Roberts and S.~A.~Hartnoll,
  {\em Pseudogap and time reversal breaking in a holographic superconductor},
  JHEP {\bf 0808}, 035 (2008)
  [arXiv:0805.3898 [hep-th]].






\bibitem{Basu:2008bh}
  P.~Basu, J.~He, A.~Mukherjee and H.~H.~Shieh,
  {\em Superconductivity from D3/D7: Holographic Pion Superfluid},
  JHEP {\bf 0911}, 070 (2009)
  [arXiv:0810.3970 [hep-th]].





\bibitem{Zeng:2010fs}
  H.~B.~Zeng, W.~M.~Sun and H.~S.~Zong,
  {\em Supercurrent in p-wave Holographic Superconductor},
  Phys.\ Rev.\  D {\bf 83}, 046010 (2011)
  [arXiv:1010.5039 [hep-th]].
























\end{thebibliography}
\end{document}